# Polarization fields in GaN/AlN nanowire heterostructures studied by Off axis holography

**M den Hertog[1], R Songmuang[1] and E Monroy[2]**

[1]Institut Néel-CNRS, 25 rue des Martyrs, 38042 Grenoble cedex 9, France
[2]CEA-Grenoble, INAC / SP2M, 17 rue des Martyrs, 38054 Grenoble cedex 9, France

E-mail: martien.den-hertog@grenoble.cnrs.fr

**Abstract.** In this work, we present an off-axis holography study of GaN/AlN heterostructured nanowires grown by plasma-assisted molecular-beam epitaxy. We discuss the sample preparation of nanowire samples for electron holography and combine potential profiles obtained using holography with theoretical calculations of the projected potential in order to gain understanding of the potential distribution in these nanostructures. The effects of surface states are discussed.

## 1. Introduction

Semiconductor nanowires (NWs) have become a powerful and diverse family of functional materials for nano-electronics, optoelectronics and biotechnology. III-nitride semiconductors, with their direct band gap, flexibility for band engineering, and capability of operating in extreme environments, are promising materials for single-NW devices, namely high-resolution chemical sensing or high-spectral-contrast photodetectors. A particularity of nitride materials is the very strong polarization-related internal electric field present in heterostructures, which can be assessed through its effect on optical or electrical properties. A direct measurement of the internal electric field profile is more challenging but could in principle be realized by off-axis electron holography. Holography on nitride materials was already reported in [1,2,3,4]. In addition to the technical difficulties to measure the field distribution, these direct measurements are necessarily perturbed by the superposition of information on surface depletion due to surface states and by the effect of strain on the internal field distribution. Taking these phenomena into account is critical to understand the physical properties of such nanostructures.

In this paper we use medium resolution off-axis electron holography to measure the projected potential in GaN/AlN NW heterostructures and compare our experimental results with three-dimensional (3D) potential simulations to discriminate the 3D field distribution and the effect of surface charge. Furthermore, we describe the challenges of holography on (nitride) NW heterostructures, which have to be faced to allow quantitative mapping of the potential.

Off-axis medium resolution electron holography was performed on an FEI Titan 80-300 working at 200 kV. Plasma-assisted molecular beam epitaxy (PAMBE) is a well-established technique to synthesize nanometer-size catalyst-free GaN NWs growing along the $c$ crystallographic axis [5,6]. This growth method allows $n$-type and $p$-type doping [7], as well as band engineering by heterostructuring [8,9,10,11,12]. For this study GaN NWs containing a 10-period GaN/AlN (7 nm / 7 nm) superlattice, followed by 30 nm of GaN and 30 nm of AlN, were grown on Si(111) patterned pillars by PAMBE (see Fig. 1). Growth proceeded under N-rich conditions (Ga/N ratio ~ 0.2–0.4) at a substrate temperature ~ 790 °C. An AlN buffer layer is used to improve the vertical alignment of the NWs [13]. The NWs under consideration have a length of 1.2–1.5 µm and a diameter of 30–80 nm. The sample for electron holography was prepared using the cleavage technique [14,15], which allowed studying epitaxial NWs while they were still attached to their substrate and suspended in vacuum, without any modification (electrical or structural) of the NW.

The sample preparation is crucial to be able to obtain good holography data from NW samples. In samples with a high density of NWs, the post geometry has several advantages over conventional dispersed NW samples on a carbon grid. First of all, a very precise orientation of the NW several degrees off the zone axis, to minimize diffraction contrast, is possible as the NWs are grown epitaxially on the Si post. Second, it is very easy to find back the same single NW as the posts also serve as markers. It should be noted that for lower density NW systems, a cleaved sample of NWs on a planar substrate can be sufficient to study free-standing NWs without overlap with other NWs close to the cleaved edge [14,15]. However, as GaN NWs grown by PAMBE are very closely packed, we need to find NWs growing at the edge of the post to prevent superposition of other NWs.

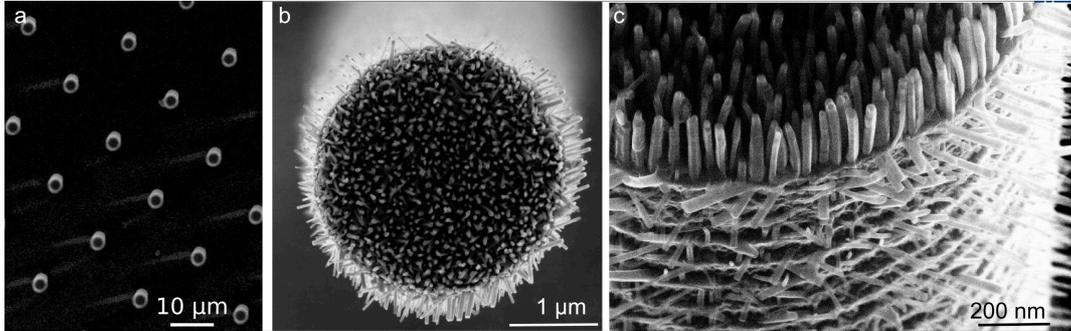

**Figure 1.** (a,b,c) Scanning electron microscopy (SEM) top view at increasing magnification of GaN/AlN NW heterostructures grown on Si (111) patterned pillars.

Previously we have demonstrated that GaN NWs are hexahedral structures with relatively flat {10-10} *m*-plane sidewall facets, and they exhibit N-polarity, as identified by annular bright field (ABF-) and high angle annular dark field scanning transmission electron microscopy (HAADF-STEM) [16]. As off-axis holography is sensitive to the projected potential we expect to observe the effect of the polarity on the potential in the structure. Figure 2(a) presents a high angle annular dark field (HAADF) STEM image of a GaN/AlN heterostructured NW. The axial superlattice is enveloped by an AlN shell arising from the lateral growth during the deposition of the AlN sections. As a result of the elevated growth temperature, a certain interdiffusion is observed at the interfaces. This phenomenon is particularly important when starting the growth of AlN on GaN, giving rise to a graded interface, which extends along about 1.5 nm. Figure 2b shows the phase image obtained by off-axis electron holography, tilted several degrees off the Si [112] zone axis (ZA). The fringe spacing was 1 nm, giving a spatial resolution in the reconstructed phase image of around 3 nm. The various GaN and AlN regions can be clearly resolved. In the absence of magnetic fields and diffraction contrast, the phase difference Δϕ between electrons that have passed through vacuum and electrons that have traversed the specimen is related to the potential $V(x, y, z)$, by

$$\Delta\varphi = C_E \int_0^t V(x,y,z)dz, \qquad (1)$$

where $C_E$ is an interaction constant which depends on the acceleration voltage ($7.29\times10^6$ rad V$^{-1}$ m$^{-1}$ at 200 kV), and $t$ is the thickness of the sample in the beam direction $z$. $V(x, y, z)$ is a combination of the mean inner potential (MIP) of the pure components, namely GaN and AlN (MIP = 16.8 V and 14.2 V, respectively [17]), and of the electrostatic potential, related to the electric field, reflecting the combined effects of spontaneous and piezoelectric polarization and surface charges. In Fig 2(c), a projected potential profile is obtained from the phase image in Fig. 2(b) along the white vertical arrow using eq. 1 and the measured NW thickness.

The experimental potential profile is compared to two simulated potential profiles obtained from projected 3D simulations of a NW structure with the layer sequence of the sample in Fig. 2(a) using the nextnano3 software [18] with the material parameters in ref. [19], one including a negative surface charge density of $-2 \times 10^{13}$ cm$^{-2}$, which is one order of magnitude higher than a value from literature [20], and one without surface charge. These simulations take into account the NW hexagonal cross section with *m* sidewall facets, with the dimensions extracted from our STEM measurements, including the presence of the AlN shell. The strain distribution is calculated to minimize the elastic energy in the system considering a free-standing NW in vacuum. The calculation of the 3D potential profile takes into consideration the local variations of the band gap due to strain, as well as the 3D internal electric field induced by spontaneous and piezoelectric polarization. However, our simulations do not include the variation in MIP between GaN and AlN; thus, the experimentally obtained potential profile was shifted along the y-axis to allow comparison to the simulated profiles. We verified that the step in phase over the AlN top region is consistent with the thickness and the MIP of AlN.

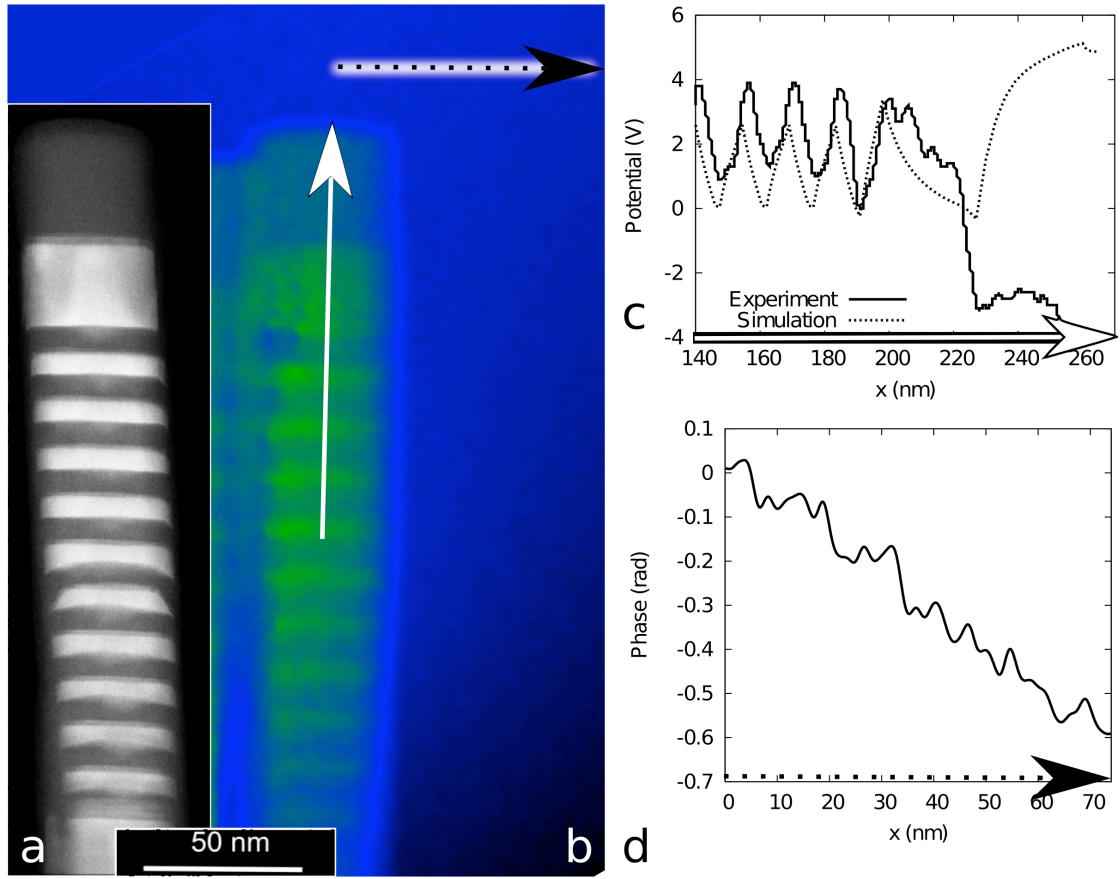

**Figure 2.** (a) HAADF STEM image of a GaN/AlN NW heterostructure. (b) Phase image of the same NW. (c) Potential profile obtained from the phase along the vertical white arrow compared with two simulated profiles one including a surface charge density of $-2 \times 10^{13}$ electron charges cm$^{-2}$ and one without surface charges. (d) Phase profile in the vacuum obtained along the horizontal black dotted arrow.

For the studied NW, the expected change in potential when going from a GaN to AlN region due only to the variation in MIP would be around 0.8 V, if we assume the NW diameter of 44 nm is roughly equal to the thickness. The experimentally measured potential change at the GaN/AlN heterointerfaces is much larger than the calculated 0.8 V, indicating that indeed the measurement is sensitive to the electric field. However, in the studied NWs there are composition gradients at the interfaces, creating also a gradient in the MIP. Therefore, we cannot provide a quantitative assessment of the interface charge in this structure, but quantitative measurements are feasible in structures with abrupt interfaces and homogeneous alloy composition.

Making a qualitative analysis, the experimental profile is comparable to the simulated profiles and consistent with the N-polarity of the NW. Interestingly, the internal field is so strong that the surface charge does not have much influence on the projected potential, even though the value of the surface charge used in the simulation is higher than experimentally measured [20]. On the other hand, the measurement at the AlN top region presents a deviation from simulations, which might be due to charging of the NW under the electron beam. To investigate charging issues, we display in Fig. 2d a phase profile in the vacuum at the side of the NW along the black horizontal dotted arrow in Fig 2b. Going away from the NW the phase decreases, indicating a positive charging of the NW [21]. This could explain the lowering of the potential in the AlN top region.

In summary we have shown first results of holography on GaN/AlN NW heterostructures, compared to theoretical calculations. We confirm the N-polarity of the NW and observe a positive charging of the sample under the electron beam. Comparison with simulated profiles does not allow setting a value on the surface charge of the NW. We have described some of the challenges that have to be met to allow quantitative field mapping in nitride NW heterostructures.

Financial support from the French FMN-SMINGUE 2011, the French CNRS and CEA METSA network, the ANR-2011-NANO-027 "UVLamp" project, the ANR 12 JS10 00201 "COSMOS" project, the EU ERC-StG "TERAGAN" (#278428) project, and the Société Française des Microscopies (SFµ) for a travel grant is acknowledged. We thank Adeline Grenier for providing the Si(111) patterned pillars as the growth substrate.